\documentclass{article}
\usepackage[english]{babel}
\usepackage[utf8]{inputenc}
\usepackage{amssymb}
\usepackage{amsmath,amsthm}
\newtheorem{theorem}{Theorem}
\usepackage{graphicx,calc}
\usepackage{subfigure}
\usepackage{tabls}

\newlength\savewidth

\title{\textbf{Fault Attacks on RSA Public Keys: \textit{Left-To-Right} Implementations are also Vulnerable}}
\author{
Alexandre Berzati
\thanks{CEA-LETI/MINATEC, 17 rue des
Martyrs, 38054 Grenoble Cedex 9, France,
\texttt{$\lbrace$alexandre.berzati,cecile.canovas$\rbrace$@cea.fr}}
\thanks{Versailles Saint-Quentin University, 45 Avenue des
  \'Etats-Unis, 78035 Versailles Cedex, France
\texttt{Louis.Goubin@prism.uvsq.fr}}
\and C\'ecile Canovas\footnotemark[1], 
\and Jean-Guillaume Dumas 
\thanks{Laboratoire J. Kuntzmann,
Universit\'e J. Fourier.
51, rue des Math\'ematiques,
umr CNRS 5224, bp 53X,
F38041 Grenoble, France,
\texttt{Jean-Guillaume.Dumas@imag.fr}
}
\and 
Louis Goubin
\footnotemark[2]
}

\begin{document}
\maketitle

\begin{abstract}
After attacking the RSA by injecting fault and corresponding
countermeasures, works appear now about the need for protecting RSA
public elements against fault attacks. We provide here an extension of
a recent attack~\cite{77} based on the public modulus corruption. The
difficulty to decompose the \textit{"Left-To-Right"} exponentiation
into partial multiplications is overcome by modifying the public
modulus to a number with known factorization. This fault model is
justified here by a complete study of faulty prime numbers with a
fixed size. The good success rate of this attack combined with its
practicability raises the question of using faults for changing
algebraic properties of finite field based cryptosystems.
 \vspace{10pt}\\
 \textbf{Keywords:} RSA, fault attacks, \textit{"Left-To-Right"} exponentiation, number theory.

\end{abstract}

\section{Introduction}
Injecting faults during the execution of cryptographic algorithms is a powerful way to recover secret information.
Such a principle was first published by Bellcore researchers~\cite{20,21} against multiple public key cryptosystems. Indeed, these papers provide successful applications including RSA in both standard and CRT modes.
This work was completed, and named Differential Fault Analysis (DFA), by E. Biham and A. Shamir with applications to secret key cryptosystems~\cite{22}. The growing popularity of this kind of attack, in the last decade, was based on the ease for modifying the behavior of an execution~\cite{12} and the difficulty for elaborating efficient countermeasures~\cite{40,43,32}.\\
\indent
Many applications against the RSA cryptosystem, based on fault
injection, have been published. The first ones dealt with the perturbation
of the private key or temporary values during the computation
\cite{20,19,21}. The perturbation of public elements  was considered
as a real threat when J-P. Seifert published an attack on the RSA
signature check mechanism~\cite{67,13}. This paper first mentions the
possibility of modifying the public modulus $N$ such that the faulty
one is prime or easy to factor. Then, E. Brier \textit{et al.}
extended this work to the full recovery of the private exponent $d$
for various RSA implementations~\cite{6}. Both works are based on the
assumption that the fault occurs before performing the RSA modular
exponentiation. A. Berzati \textit{et al.} first address the issue of
modifying the modulus during the exponentiation~\cite{77}.Still this
work was limited to an application against \textit{"Right-To-Left"}
type exponentiation algorithms.\\
\indent
In this paper we aim to generalize the previous attack to
\textit{"Left-To-Right"} type exponentiations. Under the fault
assumption that the modulus can become a number with a known
factorization, we prove that it is possible to recover the whole
private exponent. We provide a detailed study of this fault model,
based on number theory, to show its consistency and its practicability
for various kinds of perturbation. Finally, we propose an algorithm to
recover the whole private exponent that is efficient either in terms
of fault number or in computational time. 

\section{Background}
\subsection{Notations}
Let $N$, the public modulus, be the product of two large prime numbers
$p$ and $q$. The length of $N$ is denoted by $n$. Let $e$ be the
public exponent, coprime to $\varphi(N)=(p-1) \cdot (q-1)$, where
$\varphi(\cdot)$ denotes Euler's totient function. The public key
exponent $e$ is linked to the private exponent $d$ by the equation 
$e \cdot d \equiv 1 \mbox{ mod } \varphi(N)$. The private exponent $d$ is
used to perform the following operations.
\begin{description}
 \item[RSA Decryption: ]Decrypting a ciphertext $C$ boils down to compute $\tilde{m} \equiv C^{d}\mbox{ mod }N 
 \equiv C^{\sum_{i=0}^{n-1} 2^{i} \cdot d_{i}}\mbox{ mod }N$ where $d_{i}$ stands for the $i$-th bit of $d$.
 If no error occurs during computation, transmission or decryption of $C$, then $\tilde{m}$ equals $m$.
 \item[RSA Signature: ]The signature of a message $m$ is given by $S \equiv \dot{m}^{d} \mbox{ mod } N$
		       where $\dot{m} = \mu(m)$ for some hash and/or deterministic padding function $\mu$.\\
		       The signature $S$ is validated by checking that $S^{e} \equiv \dot{m} \mbox{ mod } N$.
\end{description}

\subsection{Modular exponentiation algorithms}
\begin{table}[ht]
\label{tab:l2r}
\begin{center}
\begin{tabular}{p{5.7cm}p{0.5cm}p{5.7cm}}
\cline{1-1} \cline{3-3}
\textbf{Algorithm 1:} \textit{"Right-To-Left"} modular exponentiation 
& &
\textbf{Algorithm 2:} \textit{"Left-To-Right"} modular exponentiation
\\
\cline{1-1} \cline{3-3} INPUT: $m, d, N$ & & INPUT: $m, d, N$ \\
OUTPUT: $A \equiv m^d \mbox{ mod }N$ & & OUTPUT: $A \equiv m^d \mbox{ mod }N$ \\
\cline{1-1} \cline{3-3} 1~: $A \mbox{:=} 1$; & & 1~: $A \mbox{:=} 1$;\\
2~: $B \mbox{:=} m$; & & 2~: \textbf{for} $i$ \textbf{from} $(n-1)$ \textbf{downto} $0$\\
3~: \textbf{for} $i$ \textbf{from} $0$ \textbf{upto} $(n-1)$ & & 3~: ~~~~$A \mbox{ := } A^{2}\mbox{ mod }N$;\\
4~: ~~~~\textbf{if} ($d_{i}$ == 1) & & 4~: ~~~~\textbf{if} ($d_{i}$ == 1)\\
5~: ~~~~~~~~$A \mbox{ := } (A \cdot B)\mbox{ mod }N$; & & 5~: ~~~~~~~~$A \mbox{ := } (A \cdot m)\mbox{ mod }N$;\\
6~: ~~~~\textbf{end if} & & 6~: ~~~~\textbf{end if}\\
7~: ~~~~$B \mbox{ := } B^{2}\mbox{ mod }N$; & & 7~: \textbf{end for}\\
8~: \textbf{end for} & & 8~: \textbf{return} $A$;\\
9~: \textbf{return} $A$;\\
\cline{1-1} \cline{3-3} 
\end{tabular}
\end{center}
\end{table}

Binary exponentiation algorithms are often used for computing the RSA modular exponentiation $\dot{m}^{d} \mbox{ mod } N$ where the exponent $d$ is expressed in a binary form as $d = \sum_{i=0}^{n-1} 2^{i} \cdot d_{i}$. Their polynomial complexity with respect to the input length make them very interesting to perform modular exponentiation.

The Algorithm 1 describes a way to compute modular exponentiations by
scanning bits of $d$ from least significant bits (LSB) to most
significant bits (MSB). That is why it is usually referred to as the \textit{"Right-To-Left"} modular exponentiation algorithm. This is that specific implementation that is attacked in \cite{77} by corrupting the public modulus of RSA.\\
\indent
The dual algorithm that implements the binary modular exponentiation is 
the \textit{"Left-To-Right"} exponentiation described in Algorithm 2. This algorithm scans bits of the exponent from MSB to LSB and is lighter than \textit{"Right-To-Left"} one in terms of memory consumption.

\section{Modification of the modulus and extension attempt}
\subsection{Previous work}
J-P. Seifert first addressed the issue of corrupting RSA public key elements \cite{67,13}. This fault attack 
aims to make a signature verification mechanism accept false signatures by modifying the value of the public modulus $N$. No information about the private exponent $d$ is revealed with this fault attack.
Its efficiency is linked to the attacker's ability to reproduce the fault model chosen for the modification of the modulus.\\
\indent
Seifert's work inspired the authors of~\cite{6} who first used the public modulus perturbation to recover the whole private key $d$. The attacker has to perform a perturbation campaign to gather a large enough number of (message, faulty signature) pairs. As in Seifert's attack, the fault on the modulus is induced before executing the exponentiation. Three methods based on the use of Chinese Remainder Theorem and the resolution of quite small discrete logarithms are proposed in~\cite{6} and~\cite{59} to recover the private exponent from the set of gathered pairs.\\
\indent
A new fault attack against \textit{"Right-To-Left"} exponentiation has been presented lately~\cite{77}.
This work completes the state-of-the-art by allowing the attacker to use other fault models for recovering the private exponent. The details of this work are presented below.

\subsection{Public key perturbation during RSA execution: case of the \textit{"Right-To-Left"} algorithm}

\subsubsection{Fault model.}
In J.P Seifert and E. Brier \textit{et al.}'s proposals~\cite{67,6} the fault is provoked before the exponentiation so that the whole execution is executed with the faulty modulus, $\hat{N}$.\\
The attack presented by A. Berzati \textit{et al.}~\cite{77} extends the fault model by allowing the attacker to inject the fault during the execution of the \textit{"Right-To-Left"} exponentiation. The modification of $N$ is supposed to be a transient random byte fault modification. It means that only one byte of $N$ is set to a random value. The value of the faulty modulus $\hat{N}$ is not known by the attacker.
However, the time location of the fault is a parameter known by the attacker and used to perform the cryptanalysis. This fault model has been chosen for its simplicity and practicability in smart card context~\cite{25,7}. Furthermore, it can be easily adapted to $16$-bit or $32$-bit architectures. 

\subsubsection{Faulty computation.}
Let $d = \sum_{i=0}^{n-1} 2^{i} \cdot d_i$ be the binary representation of $d$.  The output of a RSA signature can be written as:
\begin{equation}
 S \equiv \dot{m}^{\sum_{i=0}^{n-1} 2^{i} \cdot d_i} \mbox{ mod } N
\end{equation}
We consider that a fault has occurred $j$ steps before the end of the exponentiation, during the computation of a square. According to the fault model described, all subsequent operations are performed with a faulty modulus $\hat{N}$. We denote by $A \equiv \dot{m}^{\sum_{i=0}^{(n-j-1)} 2^{i} \cdot d_i}\mbox{ mod } N$ the internal register value and by $\hat{B}$ the result of the faulty square:
\begin{equation}
 \hat{B} \equiv \left(\dot{m}^{2^{(n-j-1)}}\mbox{ mod } N\right)^2 \mbox{ mod }\hat{N} 
\end{equation}
Hence, the faulty signature $\hat{S}$ can be written as:
\begin{eqnarray}
 \hat{S} & \equiv & A \cdot \hat{B}^{\sum_{i=(n-j)}^{n-1} 2^{[i-(n-j)]} \cdot d_i} \mbox{ mod } \hat{N}\\
	 & \equiv & [(\dot{m}^{\sum_{i=0}^{(n-j-1)} 2^{i} \cdot d_i}\mbox{ mod } N)\\ \nonumber
	 &        & \cdot (\dot{m}^{2^{(n-j-1)}}\mbox{ mod } N)^{\sum_{i=(n-j)}^{n-1} 
		    2^{[i-(n-j)+1]} \cdot d_i}]\mbox{ mod } \hat{N} 
\end{eqnarray}
From the previous expression of $\hat{S}$, one can first notice that the fault injection splits the computation into a correct (computed with $N$) and a faulty part (computed with $\hat{N}$). A part of $d$ is used during the faulty computation. This is exactly the secret exponent part that will be recovered in the following analysis.

\subsubsection{Attack principle.}
From both correct signature $S$ and faulty one $\hat{S}$ (obtained
from the same message $m$), the attacker can recover the isolated part
of the private key $d_{(1)} = \sum_{i=n-j}^{n-1} 2^{i} \cdot
{d_i}$. Indeed, he tries to find simultaneously candidate values for
the faulty modulus $\hat{N}'$ (according to the random byte fault
assumption) and for the part of the exponent $d_{(1)}'$ that
satisfies:
\begin{equation}
\label{eq:check1}
 \hat{S} \equiv \left( S \cdot \dot{m}^{-d_{(1)}'} \mbox{ mod }N \right) \cdot 
		\left(\dot{m}^{2^{(n-j-1)}} \mbox{ mod } N \right)^{2^{[1-(n-j)]} \cdot 
		d_{(1)}'} \mbox{ mod } \hat{N}'
\end{equation}
According to \cite{77}, the pair $(d_{(1)}',\hat{N}')$ that satisfies (\ref{eq:check1}) is the right one with a probability very close to $1$. Then, the subsequent secret bits will be found by repeating this attack using the knowledge of the already found most significant bits of $d$ and a signature faulted earlier in the process. In terms of fault number, the whole private key recovery requires an average of $\left( n/l \right)$ faulty signatures, where $l$ is the average number of bits recovered each time. As a consequence, this few number of required faults makes the attack both efficient and practicable.

\subsection{Application to the \textit{"Left-To-Right"} modular exponentiation}
\label{sec:l2r_mod_exp}
In this section, we try to apply the previously explained fault attack to the \textit{"Left-To-Right"} implementation of RSA. Under the same fault model, we wanted to know what does prevent an attacker from reproducing the attack against the dual implementation.\\
\indent 
We denote by $A$ the internal register value just before the modification of the modulus $N$:
\begin{equation}
 A \equiv \dot{m}^{\sum_{i=j}^{n-1} 2^{i-j} \cdot d_i} \mbox{ mod } N
\end{equation}
Hence, knowing that the first perturbed operation is a square, the faulty signature $\hat{S}$ can be written as:
\begin{eqnarray}
\label{eq:base_eq}
 \hat{S} & \equiv & \left( \left( \left(A^2 \cdot \dot{m}^{d_{j-1}} \right)^2 
		    \cdot \dot{m}^{d_{j-2}} \right)^2 \ldots \right)^2 
		    \cdot \dot{m}^{d_{0}} \mbox{ mod } \hat{N}\\ \nonumber
	 & \equiv & A^{2^{j}} \cdot \dot{m}^{\sum_{i=0}^{j-1} 2^i \cdot d_i} \mbox{ mod } \hat{N}
\end{eqnarray}

By observing (\ref{eq:base_eq}), one can notice that the perturbation
has two consequences on the faulty signature $\hat{S}$. First, it
splits the computation into a correct part (\textit{i.e}: the internal
register value $A$) and a faulty one, like for the perturbation of
the\textit{"Right-To-Left"} exponentiation~\cite{77}. The other one is
the addition of $j$ cascaded squares of the local variable $A$,
computed modulo $\hat{N}$. This added operation defeats the previous
attack on the \textit{"Right-To-Left"} exponentiation~\cite{77}
because of the difficulty to compute square roots in RSA rings.\\
\indent
Our idea for generalizing the previous attack to
\textit{"Left-To-Right"} exponentiation is to take advantage of the
modulus modification to change the algebraic properties of the RSA
ring. In other words, if $\hat{N}$ is a prime number, then it
is possible to compute square roots in polynomial time.
Moreover, it is actually 
sufficient that $\hat{N}$ is $B$-smooth with $B$ small enough to enable
an easy factorization of $\hat{N}$, then the Chinese Remainder Theorem
enables also to compute square roots in polynomial time.
We show next anyway that the number of primes $\hat{N}$ is sufficient to
provide a realistic fault model.

\section{Fault model}
According to the previous section, the square root problem can be
overcome by perturbing the modulus $N$ such that $\hat{N}$ is
prime. In this section we will study the consistency and the
practicability of such a fault model. Even though
 this model has already been adopted in Seifert's attack~\cite{13,67},
 we propose next further experimental evidences of the practicability
 of this model.
\subsection{Theoretical estimations}
\label{sec:prime_study}

Let us first estimate the number of primes with a fixed number of bits. From \cite[Theorem 1.10]{80}, we have the following bounds for the number of primes $\pi$ below a certain integer $x$:
\begin{eqnarray}
\label{eq:bounds}
\pi(x) & \geq & \frac{x}{\ln(x)} \left( 1 + \frac{1}{\ln(x)} +
  \frac{1.8}{\ln^2(x)} \right),~\mbox{for }x \geq 32299. \\ \nonumber
\pi(x) & \leq & \frac{x}{\ln(x)} \left( 1 + \frac{1}{\ln(x)} +
  \frac{2.51}{\ln^2(x)} \right),~\mbox{for }x \geq 355991.
\end{eqnarray}
Then, for numbers of exactly $t$ bits such that $t \geq 19$ bits, the
number of primes is $\pi_t = \pi(2^t)-\pi(2^{t-1})$. By using the
previous bounds (\ref{eq:bounds}), the probability that a $t$-bit
number is prime, $pr_t=\dfrac{\pi_t}{2^{t-1}}$, satisfies:
\begin{eqnarray}\small
 \label{eq:props}
\hspace{-15pt} pr_t > &\hspace{-5pt} {I\!n\!f}(t) &\hspace{-5pt} >  \frac{1.442t^5-3.690t^4+0.080t^3-22.828t^2+28.267t-10.810}{t^3(t-1)^3}\\  
 \nonumber
\hspace{-15pt} pr _t< &\hspace{-5pt} {S\!u\!p}(t) &\hspace{-5pt} < \frac{1.443t^5-3.689t^4+6.476t^3-35.619t^2+41.060t-15.073}{t^3(t-1)^3}
\end{eqnarray}
For instance, if $t = 1024$ bits:
$${I\!n\!f}(1024) \approx \frac{1}{709.477} \mbox{ and  }
{S\!u\!p}(1024) \approx \frac{1}{709.474}$$
Therefore around one $1024$-bit number out of $709$ is prime; and among the $2048$-bit numbers, more than one out of
$1419$ is prime.\\

For example, we construct the following sets ${\mathcal N_i}$ according to a random byte fault model.
In other words, if $\oplus$ is
the bit by bit exclusive OR, then\footnote[1]{For the sake of clarity
  we assume that a byte fault can take $2^8$ values. In fact, it can
  take only $2^8 - 1$. Indeed, the error can not be null otherwise the value of $N$ is unchanged and the fault can not be exploited.}:
\[ {\mathcal N_i}= \lbrace N \oplus R_{8} \cdot 2^{8i},~R_{8}=0~..~255
\rbrace 
\]
Let ${\mathcal N} = \bigcup_{i=0}^{\frac{n}{8}-1} {\mathcal N_i}$.
Then the cardinality of ${\mathcal N}$ and ${\mathcal N_i}$ are
$|{\mathcal N_i}|= 256$ and $|{\mathcal N}|= 256 \cdot \frac{n}{8}$. 

Would the set ${\mathcal N}$ be composed of randomly selected values,
then the proportion of primes in each ${\mathcal N_i}$ would follow
(\ref{eq:props}). 
% Hence, we can set $k = |{\mathcal N}|$ and compute
% the corresponding average and bounds with an approximation of
% $pr_t$. 
In order to estimate the number of faulty primes,
we suppose this is the case for ${\mathcal N_0}$ and that this
proportion is doubled in ${\mathcal N_i}$ for $i\geq1$, since all the
fault numbers in these sets are odd.
For $n=1024$, according to
(\ref{eq:props}), we can estimate $pr_{1024}$ and thus the average number of
faulty primes is $(2\cdot128-1)256/709.47 \approx 92.011$.
For $n=2048$, the average number of primes becomes
$(2\cdot256-1)256/1419.25\approx 92.172$.
Now for $n=1024$ but with errors of length $f$ bits, the number of
faulty primes should be around $(2\cdot\frac{1024}{f}-1)2^f/709.47$.
This gives approximately $11731$ faulty primes with $16$ bits errors
and more than $381$ millions with $32$ bits.

Moreover, consider the subset of ${\mathcal N}$ comprising only odd
numbers of size $n$. If the fault is on $f$ bits, then the latter is
of size $k=2^{f-1}(2\frac{n}{f}-1)$. Let $P\!k$ be the random variable expressing the expected number of primes in this set.
This variable follows a binomial law $\mathcal{B}(k,2pr_t)$. 
Then we can give the following confidence interval of primes (with $a$
and $b$ integer bounds): 
\begin{equation}\label{eq:intervalle}
\mbox{Pr}[a \leq P\!k \leq b] = \sum_{i=a}^b { k \choose i } (2pr_t)^i
(1-2pr_t)^{k-i}
\end{equation}
Equation
(\ref{eq:intervalle}) combined with the estimation of $pr_{1024}$ 
shows for instance that the number of faulty primes of $1024$ bits
with errors of size $8$ is comprised between $[50,140]$ in more than
$99.999$\% of the cases.

Obviously ${\mathcal N}$ is not a set of randomly chosen elements;
howbeit, empirical evidence shows that such sets behave quite like
random sets of elements, as shown below.

\subsection{Experimental results}
\begin{figure}[htbp]
 \centering
 \subfigure[Primes at consecutive $8$-bit distance of some RSA modulus]
           {\includegraphics[width=\textwidth*5/6]{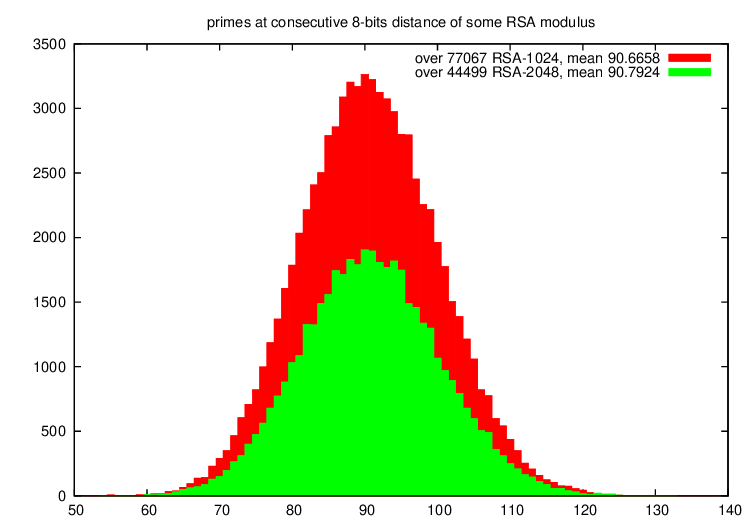}}\\
 \subfigure[Primes at consecutive $16$-bit distance of some RSA modulus]
           {\includegraphics[width=\textwidth*5/6]{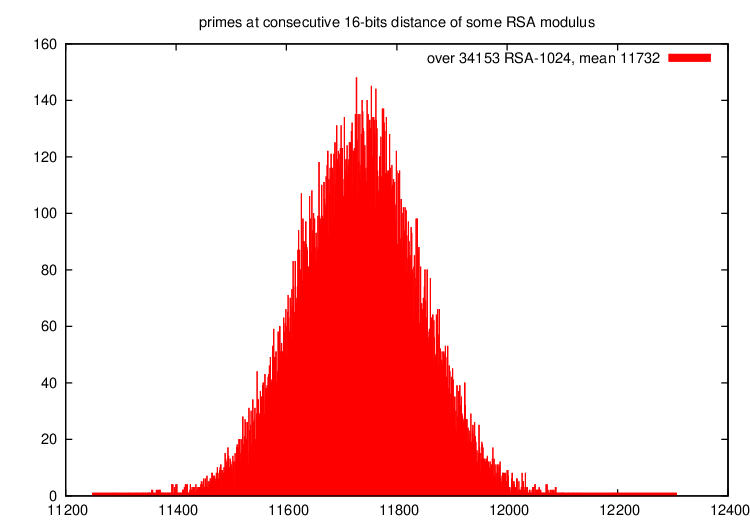}}
 \caption{Experimental distribution of primes among faulty RSA moduli}
 \label{fig:prime_rep}
\end{figure}

We have computed such sets for randomly selected RSA moduli and
counted the number of primes in those sets. The repartition seems to
follow a binomial rule (as expected) and we have the following experimental data to
support our belief (see Figure \ref{fig:prime_rep}).

As shown in Table \ref{tab:ppcount} it was anyway \textit{never} the case
that no prime was found in a set ${\mathcal N}$ (more than that we
always found more than $18$ primes in such a set). This experimental lower-bound equals to the one obtained by considering a random set. The same observation can be done for the upper-bound.
Hence, our obtained results confirm our theoretical analysis.

\begin{table}[htb]%\small
\begin{center}
\begin{tabular}{|r|r|r|r|r||r|r|r|r|}
\hline
Arch. & bits & $|{\mathcal N}|$ & Exp. \# & Prop. & \#
of & \multicolumn{3}{|c|}{\# of primes} \\
      & & &\multicolumn{2}{|c||}{of primes}&draws & Min.  & Avg.  & Max. \\
\hline
$8$-bit & 1024 & $2^{15}$ & $92.01$ & $\frac{1}{356}$ & $77067$ & $52$ & $90.6658$ & $138$\\
$8$-bit & 2048 & $2^{16}$ & $92.17$ & $\frac{1}{711}$ & $44499$ & $54$ & $90.7924$ & $132$\\
$16$-bit & 1024 & $2^{22}$ & $11731.3$ & $\frac{1}{357}$ & $34153$ & $11268$ & $11732.0$ & $12250$\\
$32$-bit & $1024$ & $2^{37}$ & $3.8\cdot 10^8$ &$\frac{1}{360}$ & & &  & \\
\hline
\end{tabular}
\caption{Experimental counts of primes in ${\mathcal N}$.}\label{tab:ppcount}
\end{center}
\end{table}

The presented results can be extended to other fault models. The Table
\ref{tab:ppcount} presents also theoretical expected results when
$16$-bit or $32$-bit architectures are targeted.
For instance, with $n=1024$ and a $16$-bit architecture, the theoretical
average number of primes 
is $11731.3$ and equation (\ref{eq:intervalle}) shows that the number
of primes should be between $[11200,12300]$ in more than $99.999$\% of the 
cases.

\subsection{Consequences} This study strengthen J-P. Seifert's assumption~\cite{67,13} of considering only prime modification of the modulus. We have showed that our fault model can be considered as a random modification of the public modulus. Then, an average of $356$ faults on $N$ will be required to obtain a prime $\hat{N}$ in the case of a $1024$-bit RSA.
\paragraph{Additional remark.}
By carefully studying the experimental results, one can notice that,
for a given modulus $N$, the byte location of the fault influences the
number of prime found in the subset. Thus, if the attacker has the
ability of setting the byte location of the fault, he can increase his
chances to get a prime faulty modulus and therefore, dramatically reduce the number of faulty signatures required to perform the attack.

\subsection{The Algorithm of Tonelli and Shanks}
\label{sec:ts}
The algorithm of Tonelli and Shanks~\cite{79} is a probabilistic and quite efficient algorithm used to compute square roots modulo $P$, where $P$ is a prime number.
The principle of the algorithm is based on the isomorphism between the multiplicative group $\left( \mathbb{Z}/P\mathbb{Z}\right)^{\ast}$ and the additive group $\mathbb{Z}/\left( P-1 \right)\mathbb{Z}$. Suppose $P-1$ is written as:
\begin{equation}
  P-1 = 2^{e} \cdot r \mbox{,~~~~~with }r \mbox{ odd}.
\end{equation}
Then, the cyclic group $G$ of order $2^{e}$ is a subgroup of $\mathbb{Z}/\left( P-1 \right)\mathbb{Z}$.
Let $z$ be a generator of $G$, if $a$ is a quadratic residue modulo $N$, then:
\begin{equation} 
 a^{\left( P-1\right)/2} \equiv \left(a^{r}\right)^{2^{e-1}} \equiv 1 \mbox{ mod } P 
\end{equation}
Noticing that $a^{r}\mbox{ mod } P$ is a square in $G$, then it exists an integer $k \in [\![0:2^{e}-1]\!]$ such that
\begin{equation}
 a^{r} \cdot z^{k} = 1 \mbox{ in } G
\end{equation}
And so, $a^{r+1} \cdot z^{k} = a \mbox{ in } G$. Hence, the square root of $a$, is given by
\begin{equation}
 a^{1/2} \equiv a^{\left( r+1\right)/2} \cdot z^{k/2} \mbox{ mod } P
\end{equation}
Both main operations of this algorithm are:
\begin{itemize}
 \item Finding the generator $z$ of the subgroup $G$,
 \item Computing the exponent $k$.
\end{itemize}
The whole complexity of this algorithm is that of finding $k$, ${\cal
  O} \left( \ln^{4} P\right)$ binary operations or ${\cal O} \left(
  \ln P\right)$ exponentiations. The details of the above algorithm
are described in~\cite{79}. In practice, on a Pentium IV $3.2$GHz, the
GIVARO\footnote[2]{GIVARO is an open source C++ library over the GNU
  Multi-Precision Library. It is available on
  \texttt{http://packages.debian.org/fr/sid/libgivaro-dev}}
implementation of this algorithm takes on average $7/1000$ of a second
to find a square root for a 1024-bit prime modulus. 

\subsection{Smooth modulus}
As in \cite{13}, what we really need for the faulty modulus is only to
be easily factorable. This is the case e.g. if the faulty $\hat{N}$
has one large prime factor and a smooth cofactor.
For instance, just testing for divisibility by the 25 first primes
(below 100), of 1024-bits faulty moduli on a $8$-bit architecture,
already roughly multiplies by $3$ the number of easily usable faulty
moduli. 
Then, one can also compute square roots modulo a
non prime modulus as long as the factorization is known. The idea is
first to find square roots modulo each prime factors of
$\hat{N}$; then to lift them independently to get square roots modulo
each prime power; and finally to combine them using the Chinese
Remainder Theorem (see e.g. \cite[\S 13.3.3]{83} for more details). 
The number of square roots increases but since they are computed on
comparatively smaller primes, the overall complexity thus remains
${\cal O} \left( \ln^{4} \hat{N}\right)$ binary operations. 
For the sake of simplicity, in the following
we consider only prime faulty moduli.

\section{Cryptanalysis}
\label{sec:crypt}
The purpose of our fault attack against the \textit{"Left-To-Right"} exponentiation is similar to the attack against the \textit{"Right-To-Left"} one~\cite{77}. The modulus $N$ is transiently modified to a prime value during a squaring, $j_k$ steps before the end of the exponentiation. Then, from a correct/faulty signature pair $(S,\hat{S}_{k})$, the attack aims to recover the part of private exponent $d_{(k)} = \sum_{i=0}^{j_{k}-1} 2^i \cdot d_i$ isolated by the fault. By referring to \cite{77}, the following analysis can be easily adapted for faults that first occurs during a multiplication.

\subsection{Dictionary of prime modulus.}
The first step consists in computing a dictionary of prime faulty modulus candidates $(\hat{N_i})$. The attacker tests all possible values obtained by modifying $N$ according to a chosen fault model. Then, candidate values for $\hat{N}$ are tested using the probabilistic Miller-Rabin algorithm~\cite{81}. According to our study (see Sect. \ref{sec:prime_study}), for a random byte fault assumption, the faulty modulus dictionary will contain 46 entries in average either for a 1024-bit or a 2048-bit RSA. The size of the dictionary depends on the fault model (see Table \ref{tab:ppcount}).
\subsection{Computation of square roots.}
\label{sec:sqroots}
For each entry $\hat{N_i}$ of the modulus dictionary, the attacker
chooses a candidate value for the searched part of the private
exponent $d_{(k)}'$. Now he can compute\footnote[3]{This computation
  is possible only when $d_{k}'$ is invertible in
  $\mathbb{Z}/\mathbb{Z}\hat{N_i}$; in our case all the considered
  $N_i$ are primes and Euclid's algorithm always computes the inverse.}:
\begin{equation}
 \label{eq:comp_r}
 R_{(d_{(k)}',\hat{N_i})} \equiv \hat{S}_{k} \cdot \dot{m}^{-d_{(k)}'} \mbox{ mod } \hat{N_i}
\end{equation}
For the right pair $(d_{(k)},\hat{N})$, $R_{(d_{(k)},\hat{N})}$ is expected to be a multiple quadratic residue (\textit{i.e:} a $j_k$-th quadratic residue, see Sect. \ref{sec:l2r_mod_exp}).
As a result, if $R_{(d_{(k)}',\hat{N_i})}$ is not a quadratic residue, the attacker can directly deduce that the candidate pair $(d_{(k)}',\hat{N_i})$ is a wrong one. The quadratic residuosity test can be done in our case because all precomputed candidate values for the faulty modulus are prime numbers. The test is based on Fermat's theorem:
\begin{eqnarray}
 \mbox{If } {\left(R_{(d_{(k)}',\hat{N_i})}\right)}^{\left(\hat{N_i}-1\right)/2 } \equiv 1 \mbox{ mod } \hat{N_i}
\end{eqnarray}
\begin{center}
then $R_{(d_{(k)}',\hat{N_i})} \mbox{is a quadratic residue modulo }\hat{N_i}$ 
\end{center}
If the test is satisfied then the attacker can use the Tonelli and Shanks algorithm (see Sect. \ref{sec:ts}) to compute the square roots of $R_{(d_{(k)}',\hat{N_i})}$. 
Therefore, to compute the $j_k$-th square root of $R_{(d_{(k)}',\hat{N_i})}$, this step is expected to be repeated $j_k$-times. But, when one of the $j_k$ quadratic residuosity test fails, the current candidate pair is directly $(d_{(k)}',\hat{N_i})$ rejected and the square root computation is aborted. The attacker has to choose another candidate pair.

\subsection{Final modular check.}
The purpose of the two first steps is to cancel the effects on the faulty signature due to the perturbation.
Now, from the $j_k$-th square root of $R_{(d_{(k)}',\hat{N_i})}$ the attacker will simulate an error-free end of execution by computing:
\begin{equation}
\label{eq:finalchk1}
 S' \equiv {\left({\left(R_{(d_{(k)}',\hat{N_i})}\right)}^{1/2^{j_k}} \mbox{ mod } \hat{N_i}\right)}^{{2}^{j_k}} 
 	   \cdot \dot{m}^{d_{(k)}'} \mbox{ mod } N
\end{equation}
Finally, he checks if the following equation is satisfied:
\begin{equation}
\label{eq:finalchk2}
 S' \equiv S \mbox{ mod } N
\end{equation}
As in the \textit{"Right-To-Left"} attack~\cite{77}, when this latter
condition is satisfied, it means that the candidate pair is very
probably the searched one (see Sect. \ref{sec:proba}). Moreover, the
knowledge of the already found least significant bits of $d$ is used
to reproduce the attack on the subsequent secret bits. As a
consequence, the attacker has to collect a set of faulty signatures
$\hat{S}_{k}$ by injecting the fault at different steps $j_k$ before
the end of the exponentiation. Moreover, multiple faulty signature
$\hat{S}_{k,f}$ have to be gathered for a given step $j_k$ to take
into account the probability for having a faulty signature
$\hat{S}_{k}$ computed under a prime $\hat{N}$, that is to say
exploitable by the cryptanalysis. 
This set $(\hat{S}_{k,f}, j_k)_{k,f}$ is sorted in descending fault
location. If faults are injected regularly, each sorted pair is used
to recover a $l$-bit part of the exponent such that for the $k$-th
pair $(\hat{S}_{k,f}, j_k)$, the recovered part of $d$  
is $d_{(k)} = {\sum}^{j_{k} -1}_{i=0} 2^i \cdot d_i = {\sum}^{k \cdot
  l -1}_{i=0} 2^i \cdot d_i$. These results can be applied for faults
that are not injected regularly (\textit{i.e}: $j_k - j_{k-1} = l_k <
l_{max}$). The attack algorithm is given in more details next.

\begin{table}[htbp]
\begin{center}
\begin{tabular}{l}
\hline
\hline
\textbf{Algorithm 3:} DFA against \textit{"Left-To-Right"} modular exponentiation\\
\hline
INPUT: $N$, $\dot{m}$, the correct signature $S$, the size of the dictionary $D_{length}$,\\
~~~~~~~~~~~~the set of pairs $(\hat{S}_{k,f}, j_k)_{0\leq k < n/l,~1\leq f \leq \mu\left( F_n \right)}$\\
OUTPUT: the private exponent $d$\\
\hline
~1: \textit{//Computation of the dictionary of prime faulty modulus candidates}\\
~2: Dict = $Build\_Prime\_Dict$($N$, $D_{length}$);\\
~3: \textit{//Initialization}\\
~4: $d$ := $0$;\\
~5: \textit{//All the faulty signatures are tested}\\
~6: \textbf{for} $k$ \textbf{from} $0$ \textbf{upto} $\lfloor n/l \rfloor$\\
~7: ~~~\textbf{for} $f$ \textbf{from} $1$ \textbf{upto} $\mu\left( F_n \right)$\\
~8: ~~~~~~\textbf{for} $d_{(k)}$ \textbf{from} $0$ \textbf{upto} $2^l-1$\\
~9: ~~~~~~~~~$d'$ := $d_{(k)} \cdot 2^{j_k} +d$;\\
10: ~~~~~~~~~\textbf{for} $i$ \textbf{from} $1$ \textbf{upto} $D_{length}$\\
11: ~~~~~~~~~~~~$R$ := $\hat{S}_{k,f} \cdot \dot{m}^{-d'} \mbox{ mod } Dict[i]$;\\
12: ~~~~~~~~~~~~\textit{//The function computes $j_k$ square roots and returns $0$ when a test fails}\\
13: ~~~~~~~~~~~~$R$ := $Test\_And\_Tonelli$($R$, $j_k$, $Dict[i]$);\\
14: ~~~~~~~~~~~~\textit{//If a test fails, then we have to test another candidate pair}\\
15: ~~~~~~~~~~~~\textbf{if} $\left( R == 0\right)$\\
16: ~~~~~~~~~~~~~~~\textbf{break};\\
17: ~~~~~~~~~~~~\textbf{else}\\
18: ~~~~~~~~~~~~~~~$S'$ := $R^{2^{j_k}} \cdot \dot{m}^{d'} \mbox{ mod } N$\\
19: ~~~~~~~~~~~~~~~\textit{//Final check}\\
20: ~~~~~~~~~~~~~~~\textbf{if} $\left( S' == S \mbox{ mod } N \right)$\\
21: ~~~~~~~~~~~~~~~~~~\textit{//The attack continues for the subsequent $l$-bit part of $d$}\\
22: ~~~~~~~~~~~~~~~~~~$d$ := $d'$;\\
23: ~~~~~~~~~~~~~~~~~~\textbf{goto} $line\_6$;\\
24: ~~~~~~~~~~~~~~~\textbf{end if};\\
25: ~~~~~~~~~~~~\textbf{end if};\\
26: ~~~~~~~~~\textbf{end for};\\
27: ~~~~~~\textbf{end for};\\
28: ~~~\textbf{end for};\\
29: \textbf{end for};\\
30: \textbf{return} $d$;\\
\hline
\end{tabular}
\end{center}
\end{table}

\section{Performance}

\subsection{Fault number}
Our fault model is based on modifications of the modulus $N$, with
a fault of size $l$ bits, such that its corresponding faulty value is prime. 
In Section \ref{sec:prime_study}, we have shown that the probability
for a $t$-bit number to be prime, $pr_t$, can be bounded. Now, let the
number of fault to make $\hat{N}$ prime be the random variable
${F_t}$. 
% This random variable follows a geometric probability
% law. 
We have seen that the average number of faults to make $\hat{N}$ prime
per faulted part is:
\begin{equation}
\dfrac{2^l}{(2\cdot 2^l -1)S\!u\!p(t)} < \mu \left( {F_t}\right) < \dfrac{2^l}{(2\cdot 2^l -1)I\!n\!f(t)} 
\end{equation}
For large values of $t$ (\textit{i.e}: at least $1024$ or $2048$-bit
RSA), we can use the pinching (or sandwich) theorem to approximate
this value asymptotically :
\begin{eqnarray}
 \mu \left( {F_t}\right) & \sim & \dfrac{t}{1.4427\cdot(2-\frac{1}{2^l})} 
\end{eqnarray}
From a given faulty signature, the attacker can recover a $l$-bit part
of $d$. There are at most $n/l$ such parts for an RSA of size $n$.
This shows that the average number of faults required for a whole
private key satisfies:
\begin{equation}
  \mbox{Number of faults} = {\cal O}\left( \dfrac{n^2}{2.8854 \cdot l}\right) \mbox{tries}
\end{equation}
This number can be dramatically reduced if the attacker has the
ability to chose the byte location of the fault (see
Sect. \ref{sec:prime_study}) or if the fault model is larger
(\textit{i.e}: smooth modulus, different architectures targeted
\ldots).

\subsection{Computational complexity}
% For the sake of clarity, we have chosen to directly express the
% computational complexity of the analysis for a whole private key
% recovery. Further details are provided in Appendix \ref{sec:cpx}.

We now give the overall complexity of the attck. 
The size of the dictionary, $D_{length}$, is let as an attack
parameter since the
attacker can fix a limit if the chosen fault model requires more
resources than he can get. 
According to our previous analysis (see Sect. \ref{sec:prime_study}),
$D_{length} = 46$ for a random byte fault assumption. 

\begin{theorem}Algorithm 3 is correct and its average complexity for a random byte fault perturbation of the modulus satisfies:
 $$ {C}_{attack} = {\cal O}\left(\dfrac{2^{8+l} \cdot n^{3} \cdot (n+l)}{16 \cdot l}\right)
\mbox{ exponentiations}
$$\end{theorem}
\begin{proof}Correctness as been shown in section \ref{sec:crypt}. 
Now for the complexity, the attacker has to test all possible
candidate pairs $(d_{(k)}',\hat{N_i})$.
The number of pairs depends on the size of the dictionary of prime
modulus denoted by $D_{length}$ and the window recovery length $l$:
\begin{equation}
|(d_{(k)}',\hat{N_i})|  = 2^{l} \cdot D_{length}
\end{equation}

For each pair the attacker first computes $R_{(d_{(k)}',\hat{N_i})}$ (see (\ref{eq:comp_r})) by executing a modular exponentiation of the message and a multiplication.\\
Then, he performs a series of at most $j_k$ 
quadratic residuosity tests and, for
each success, a square root is computed. By noticing that the
probability to fail in the test follows a geometric probability law,
the average number of performed tests\footnote[4]{The test fails when tested
  value is not a quadratic residue. But all the $\hat{N_i}$ are prime.
Let be $z_i$ a generator in $\mathbb{Z}/\hat{N_i}\mathbb{Z}$, all the
elements of the group can be expressed as a power of ${z_{i}}$. Hence
one element out of 2 is a power of ${z_{i}}^2$ and a quadratic
residue.} is $\frac{1}{ \mbox{Pr[Test fails]}} = 2$.
 As a consequence, the average complexity of this step
is:
\begin{eqnarray}
 {C}_{Square~roots}(k) & = & {\cal O}\left( 2 \cdot {C}_{Test} + {C}_{Tonelli~\&~Shanks} \right)\\\nonumber
		    & = & {\cal O}\left(j_k \cdot n \right) \mbox{ exponentiations} 
\end{eqnarray}
The last step of the attack is the final check (see
(\ref{eq:finalchk1})). It requires to compute $j_k$ modular squares
and a modular exponentiation of the message followed by a
multiplication. The latter computation is also bounded by 
${\cal O}(j_k \cdot n)$ exponentiations. 

Now in the case of a fixed size dictionary the average number of
primes of this dictionary for a byte modification of the modulus is
${N}_{faults~per~blocs} = \frac{2^8 n/8}{D_{length}}$.

Then, the  attack has to test all of the
gathered faulty signatures in order to recover the whole exponent.
Hence, as $j_k$ is bounded by $k \cdot l$, the overall
computational complexity is bounded by:
\begin{eqnarray}
  {C}_{attack} & = & \sum_{k=0}^{n/l} {N}_{faults~per~blocs} \cdot
  {C}_{Square~roots}(k) \cdot 2^{l} \cdot D_{length}\\\nonumber 
& = & {\cal O}\left(\dfrac{2^{8+l} \cdot n^{3} \cdot (n+l)}{16 \cdot l}\right)
\end{eqnarray}
\end{proof}

The presented attack is thus longer than the
\textit{"Right-To-Left"} one~\cite{77}, the principal reason being
the extra number of faulty pairs to analyze in order to get a prime modulus.

\subsection{False-acceptance probability}
\label{sec:proba}
As defined in~\cite{77}, the false-acceptance probability is the probability for a wrong pair $(d_{(k)}',\hat{N_i})$ to satisfy (\ref{eq:finalchk2}). In our case, the computation of the final check is done in $\mathbb{Z}/N\mathbb{Z}$ and requires extra squares. As a consequence the false-acceptance probability given in~\cite{77} has to be adapted by replacing the search space for $\hat{N}$ by the dictionary length $D_{length}$: 
\begin{equation}
 0 < \mbox{Pr}[F.A] <	min\left(\frac{(N\!-\!1)\!\cdot\!2^{l}\!\cdot\!D_{length}}
				      {N\!\cdot\!(2^{l}\!\cdot\!D_{length}-1)},\\
			\frac{2^{l}\!\cdot\!D_{length}}{N}\right)
\end{equation}
Moreover, because of the quadratic residuosity tests (see Sect. \ref{sec:sqroots}), false candidates can be rejected before computing the final check. Hence, the final check will not always be done. The probability that a wrong pair pass all the $j_k$ tests is given by:
\begin{eqnarray}
 &   & \mbox{Pr}\left[ R_{(d_{(k)}',\hat{N_i})} \mbox{is a } j_k\mbox{-times quadratic residue}\right]\\
 \nonumber
 & = & \prod_{i=0}^{j_k-1}\mbox{Pr} \left[ {\left(R_{(d_{(k)}',\hat{N_i})}\right)}^{1/2^i}
					     \mbox{is a quadratic residue} \right]\\ \nonumber
 & = & \frac{1}{2^{j_k}} 
\end{eqnarray}
This probability indicates that, for recovering the $k$-th part of $d$, only one out of $2^{j_k}$ wrong pairs will pass all the quadratic residuosity tests. Eventually, the false-acceptance probability can be upper-bounded:
\begin{equation}
     \mbox{Pr}[F.A] < min \left(\frac{1}{2^{j_k}}, \frac{(N\!-\!1)\!\cdot\!2^{l}\!\cdot\!D_{length}}
				      {N\!\cdot\!(2^{l}\!\cdot\!D_{length}-1)},\\
			\frac{2^{l}\!\cdot\!D_{length}}{N}\right)
\end{equation}
This expression first shows that because of the last term $\frac{2^{l}\!\cdot\!D_{length}}{N}$, the false-acceptance probability is highly negligible for commonly used RSA length. Furthermore, 
one can advantageously notice that the final check can be avoided when the number of consecutive quadratic residuosity tests to pass is large enough (\textit{i.e}: $2^{j_{k}} > D_{length} \cdot 2^{l}$).

\section{Conclusion}
In this paper, we generalize the fault attack presented in~\cite{77} to \textit{"Left-To-Right"} implementation of RSA by assuming that the faulty modulus can be prime. Although this model has been already used~\cite{67}, this  paper provides a detailed theoretical analysis in fault attack context. Furthermore this analysis proves that such a fault model is not only practicable but extendible to different architectures. This emphases the need for protecting RSA public elements during the execution.\\
\indent
More generally the use of a faulty prime modulus to compute square
roots in polynomial time raises the question of using faults for
changing algebraic properties of the underlying finite domain. 
This paper provides an element of answer that may be completed by
future fault exploitations.

\section*{Acknowledgment}
We would like to thank Sonia Belaid and Louis Schimchowitsch for
finding an error in the faulty primes search program, in an earlier
version of this report.
%References
\bibliographystyle{alpha}
\bibliography{biblio}

\end{document}